\documentclass[]{pasj02} 
\usepackage[switch,mathlines]{lineno} % add line number to manuscript

\jyear{2025}
\Received{2025/5/12}%{yyyy/mm/dd}
\Accepted{2025/8/05}%{yyyy/mm/dd}
\graphicspath{{./}{figures/}} 

\usepackage{comment}
\usepackage{ulem}
\usepackage{url}
\usepackage{multirow}
\usepackage[legacycolonsymbols]{mathtools} 
\usepackage{booktabs}

\begin{document} 

\title{Comparison of simulations and semi-analytical model for WDM subhalo mass functions}

%%% begin:list of authors
% Do NOT capitalize all letters in "textsc".
\author{
Mizuki \textsc{Ono},\altaffilmark{1}\altemailmark \email{mizuki@astro1.sci.hokudai.ac.jp} 
 Takashi \textsc{Okamoto},\altaffilmark{2}\altemailmark\orcid{0000-0003-0137-2490} \email{takashi.okamoto@sci.hokudai.ac.jp} 
 Shin'ichiro \textsc{Ando},\altaffilmark{3,4}\orcid{0000-0001-6231-7693}
 and 
 Tomoaki \textsc{Ishiyama}\altaffilmark{5}\orcid{0000-0002-5316-9171}
}
\altaffiltext{1}{Department of Cosmosciences, Graduate School of Science, Hokkaido University, N10 W8, Kitaku, Sapporo 060-0810, Japan}
\altaffiltext{2}{Department of Physics, Faculty of Science, Hokkaido University, N10 W8, Kitaku, Sapporo 060-0810, Japan}
\altaffiltext{3}{GRAPPA Institute, University of Amsterdam, Science Park 904, 1098 XH Amsterdam, The Netherlands}
\altaffiltext{4}{Kavli Institute for the Physics and Mathematics of the Universe, University of Tokyo, Chiba 277-8583, Japan}
\altaffiltext{5}{Institute of Management and Information Technologies, Chiba University, 1-33, Yayoi-cho, Inage-ku, Chiba 263-8522, Japan}

%\footnotetext[$\dag$]{Present address: ....}

%%% end:list of authors

%% !!! Select 3 to 5 words from PASJ's key words !!! 
%% List of Key Words: https://academic.oup.com/pasj/pages/Pasj_Keywords 
%% "\KeyWords{ }" always has to be placed before ``\maketitle'' 

\maketitle

\begin{abstract}
The Cold Dark Matter (CDM) model successfully explains large-scale structure formation, but challenges remain at smaller scales, leading to interest in Warm Dark Matter (WDM) as an alternative. The abundance of Milky Way subhalos depends on the mass of WDM particles, allowing constraints to be obtained by comparing observations and theoretical models. However, high-resolution simulations of heavier WDM particle masses are computationally demanding, making semi-analytical approaches valuable. In this study, we evaluate the ability of the Semi-Analytical Sub-Halo Inference ModelIng for WDM (SASHIMI-W) to reproduce subhalo mass functions for heavier WDM particle masses. We perform high-resolution cosmological N-body simulations for CDM and WDM with particle masses of 1 keV, 3 keV, and 10 keV, and compare the ratio of the subhalo mass function between WDM and CDM cases. Our results show that SASHIMI-W successfully reproduces the simulation results over redshifts z = 0 to z = 2. Furthermore, both simulations and the semi-analytical model show a slight redshift dependence in the subhalo suppression ratio. However, a direct comparison of the differential subhalo mass functions shows discrepancies in the mid- and low-mass regions, suggesting that the tidal stripping effects implemented in SASHIMI-W may be too strong for WDM subhalos, or that the removal of spurious subhalos in the simulations is insufficient. These results validate the use of SASHIMI-W in constraining WDM properties, and highlight the need for refinements in both tidal effect modeling and spurious subhalo filtering to improve subhalo abundance predictions.
\end{abstract}

%\pagewiselinenumbers 

\section{Introduction}
The true nature of dark matter is still unclear. However, its existence has been confirmed by observational evidence. Observations of the Cosmic Microwave Background (CMB) suggest that dark matter consists of non-baryonic particles (e.g., \cite{2011ApJS..192...16L}), although the specific type of particle remains unknown. At present, Cold Dark Matter (CDM), which has negligible thermal velocity at the epoch of decoupling, has gained significant attention, with axions being a strong candidate (\cite{1983PhLB..120..127P}). Indeed, the CDM model explains observations on large scales very well, but several problems have been pointed out on smaller scales, such as the missing satellite problem \citep{1999ApJ...522...82K,1999ApJ...524L..19M}, the core-cusp problem \citep{1994ApJ...427L...1F,1994Natur.370..629M,10.1093/mnras/stu972}, and the too-big-to-fail problem \citep{Boylan_Kolchin_2011,10.1111/j.1365-2966.2011.19971.x}. On such small scales, the distribution of dark matter is strongly influenced by baryons, which means that the properties of dark matter on these scales are not well constrained. Therefore, there is room to consider dark matter that behaves like CDM on large scales but differently on small scales, with warm dark matter (WDM) being one candidate.

Unlike CDM, WDM particles possess thermal velocity in the early universe, with sterile neutrinos being a strong candidate (e.g., \cite{Asaka_2005}). When WDM particles decouple from the primordial plasma, they remain relativistic, allowing them to stream freely and erase small-scale density fluctuations. As the universe expands, WDM particles gradually lose energy and transition to non-relativistic particles during the radiation-dominated era. By the time this transition occurs, WDM particles have eliminated density fluctuations with wavelengths shorter than their free-streaming scale. This introduces a characteristic cutoff in the linear matter power spectrum, suppressing structure formation below this scale.
Furthermore, when WDM particles accumulate at the center of a dark matter halo, their non-negligible thermal velocity results in a lower phase-space density compared to CDM. This causes the central density distribution to flatten into a core-like structure with nearly constant density (\cite{PhysRevD.62.063511}). However, theoretical models suggest that cores formed through this process are significantly smaller than those observed in the Milky Way's satellite dwarf galaxies (\cite{2011JCAP...03..024V}), indicating that additional physical mechanisms may be necessary to explain the observed core sizes.

WDM models have the potential to address some of the shortcomings of the CDM paradigm, including the missing satellite and too-big-to-fail problems, thanks to its intrinsic characteristics. However, WDM also presents certain challenges, including the lack of a well-defined mass range. The lower bound on the mass of the WDM particle is $5-6\ \rm keV$, as determined by strong gravitational lensing observations \citep{2020MNRAS.491.6077G,10.1093/mnras/sty2393,10.1093/mnras/stz3177}. Similarly, Lyman-$\alpha$ forest observations have set a lower bound of $3.5\ \rm keV$ \citep{annurev:/content/journals/10.1146/annurev-astro-082214-122355,PhysRevD.96.023522,10.1093/mnras/stab1960}. However, it should be noted that these constraints are dependent on the observational methods employed. 

Investigating the number of satellite galaxies in the Milky Way offers a promising way to constrain the WDM particle mass. WDM models predict fewer subhalos within the dark matter halo compared with CDM, with the degree of suppression depending on the particle mass \citep{Lovell_2014}.
An accurate theoretical model of the subhalo mass function would allow the WDM particle mass to be constrained by comparison with observational data. 
Cosmological $N$-body simulations have proven their ability to estimate subhalo properties \citep{Lovell_2014,Tormen_1997,10.1111/j.1365-2966.2004.08360.x}. 

However, cosmological simulations face significant computational challenges, particularly for colder (i.e. higher mass) WDM models.
The main difficulty lies in resolving density perturbations whose wavelengths are shorter than the free-streaming length, which is essential for accurately capturing the suppression of small-scale structure formation. 
As the WDM particle mass increases, the free-streaming length decreases, requiring higher spatial and mass resolution to resolve the relevant perturbations. Consequently, the computational cost increases sharply, making simulations prohibitively expensive.

Analytical and semi-analytical models have become more computationally efficient alternatives.
Prior studies employed rudimentary semi-analytical models based on the extended Press-Schechter (EPS) theory to assess WDM model viability \citep{10.1093/mnras/stv1169, Cherry_2017}. A key limitation of these approaches was the lack of comprehensive mass evolution models for subhalos, particularly critical factors such as tidal effects that significantly impact subhalo dynamics and structure.

\citet{2022PhRvD.106l3026D} developed the Semi-Analytical Sub-Halo Inference ModelIng for WDM (SASHIMI-W), which provided a framework for modeling mass evolution in warm dark matter (WDM) cosmologies \citep{2022PhRvD.106l3026D,2018PhRvD..97l3002H,2016MNRAS.460.1214L}. SASHIMI-W is based on the cold dark matter (CDM) version of SASHIMI (SASHIMI-C; \cite{2018PhRvD..97l3002H}, \cite{ando2019}, \cite{2022MNRAS.517.2728H}) and has been extended for WDM by modifying the mass-loss rate and implementing appropriate changes to the extended Press–Schechter (EPS) formalism, as well as to the relationships between concentration, mass, and redshift.

They used SASHIMI-W to test WDM models for Milky Way-mass halos within the mass range $0.6\times 10^{12}\, \mathrm{M}_\odot$ to $2.0\times10^{12}\, \mathrm{M}_\odot$, excluding WDM models with masses less than $3.6-5.1\ \rm keV$ at the $95\%$ confidence interval. However, their comparison relied on the fitting function from \citet{Lovell_2014}, which was derived from simulations of WDM models with masses between $1.5-2.3\ \rm keV$. Consequently, this study seeks to evaluate SASHIMI-W's ability to reproduce simulations of WDM models with higher particle masses.

In this study, we perform cosmological $N$-body simulations of CDM and WDM models with WDM masses of 1 keV, 3 keV, and 10 keV for halos with masses comparable to the Milky Way at redshift $z=0$. We then compare the ratio $r$ of the mass functions of the WDM and CDM subhalos obtained from the simulations with those obtained from SASHIMI-W.
The ratio $r$ is given by
\begin{equation}
    r = \dfrac{n_{\rm WDM}}{n_{\rm CDM}} = \dfrac{{\rm d}N_{\rm WDM}/{\rm d}\log{M_{\rm sub}}}{{\rm d}N_{\rm CDM}/{\rm d}\log{M_{\rm sub}}},
    \label{r}
\end{equation}
where $n_{\rm X}$ is the differential mass function of subhalos in a model X, $M_{\rm sub}$ is the mass of a subhalo, and ${\rm d}N_{\rm X}$ is the number of subhalos in a given mass range in the model X.

This paper is organized as follows. In Section 2, we describe our simulations and semi-analytic model (SASHIMI-W). In Section 3, we present the key results of our analysis, examining the model's performance. In Section 4, we discuss the potential sources of discrepancy between the simulation results and the SASHIMI-W predictions. Finally, in Section 5, we summarize our conclusions.

\section{Simulations and the semi-analytical model}\label{sec:2}
This section details our computational methodology. We begin by presenting the simulation setup, followed by our approach to mitigating numerical artifacts through the removal of spurious subhalos. We then briefly introduce SASHIMI-W, our semi-analytic model.

\subsection{Simulations}
We perform cosmological $N$-body simulations of halos with a mass of $\sim 10^{12}\, \mathrm{M}_\odot$ at $z=0$, spanning CDM models and WDM models with particle masses of 1 keV, 3 keV, and 10 keV. Two distinct simulation sets are employed: the first set includes 1 keV and 10 keV WDM models along with a CDM. The zoom-in simulations target a halo with a mass of approximately $10^{12}\, \mathrm{M}_\odot$ at $z=0$, corresponding to a Milky Way-mass halo. The second set comprises the WDM model of 3 keV and a CDM model (CDM2), using uniform resolution initial conditions. The cosmological parameters are based on the Planck 2018 results \citep{2020A&A...641A...6P} and on the Planck 2016 results \citep{2016A&A...594A..13P} for the second set. 
The specific values used in each case are summarized in Table \ref{tab:cosmo}.
%Table 1
\begin{table}[h]
    \tbl{Cosmological parameters}{
    \begin{tabular}{@{}l|ccccc@{}}
    \hline
       Parameter & $\Omega_{\rm m}$ & $\Omega_{\rm \Lambda}$ & $h$ & $n_{\rm s}$ & $\sigma_8$\\ \hline
       Zoom-in & $0.3158$ & $0.6842$ & $0.6732$ & $0.96606$ & $0.81$\\ 
       Uniform box & $0.31$ & $0.69$ & $0.68$ & $0.96$ & $0.83$\\ \hline 
    \end{tabular}}\label{tab:cosmo}
\end{table}

The mass of a single dark matter particle, $m_\mathrm{p}$, is $2.5\times 10^3 \, h^{-1} \, \mathrm{M}_\odot$ in the first set of simulations and $4.1\times 10^4 \, h^{-1} \, \mathrm{M}_\odot$ in the second set. The mean interparticle spacing, $d$, is $3.05\times 10^{-3} \, h^{-1} \, \mathrm{Mpc}$ for the first set and $7.81\times 10^{-3} \, h^{-1} \, \mathrm{Mpc}$ for the second set. These simulation parameters are summarized in Table~\ref{tab:simparam}.

In the first set (zoom-in simulations), a comoving gravitational softening length of $8.415\times 10^{-2} \, h^{-1} \, \mathrm{pc}$ is applied up to redshift $z = 9$. Below this redshift, the softening length is fixed at $8.415\times 10^{-3} \, h^{-1} \, \mathrm{pc}$ in physical units. In contrast, the second set (uniform-resolution simulations) employs a constant comoving softening length of $120 \, h^{-1} \, \mathrm{pc}$ for dark matter particles throughout the entire simulation.
%Table 2
\begin{table*}[h]
    \tbl{Simulation parameters. Here, $m_p$, $d$, and $L$  are the mass of a single dark matter particle, the mean interparticle spacing, and the box size, respectively.}{
        \begin{tabular}{@{}l|ccccc@{}}
        \hline
        Parameter & 1 keV & 3 keV & 10 keV & CDM1 & CDM2 \\
        \hline
        $m_p [{\rm M}_{\odot}/h]$ & $2.5\times 10^3$ & $4.1 \times 10^4$ & $2.5 \times 10^3$ & $2.5 \times 10^3$ & $4.1\times 10^4$ \\
        $d [{\rm Mpc}/h]$ & $3.05\times 10^{-3}$ & $7.81\times 10^{-3}$ & $3.05\times 10^{-3}$ & $3.05\times 10^{-3}$ & $7.81\times 10^{-3}$ \\
        $L [{\rm Mpc}/h]$ & $25$ & $8$ & $25$ & $25$ & $8$ \\ \hline
        \end{tabular}
    }\label{tab:simparam}
\end{table*}

We run simulations over a redshift range from $z = 127$ to $z = 0$. The simulations in the first set are conducted using GIZMO \citep{2015MNRAS.450...53H}, while those in the second set are performed with GreeM \citep{2009PASJ...61.1319I}. 
The initial conditions are generated using MUSIC (\cite{2011MNRAS.415.2101H}) for the zoom-in simulation and, for the uniform box, they are generated with 2LPTIC (\cite{2006MNRAS.373..369C}). 
Furthermore, dark matter halos and subhalos are identified by using the ROCKSTAR halo finder (\cite{2013ApJ...762..109B}), and merger trees are constructed with CONSISTENT-TREES (\cite{2013ApJ...763...18B}).

In WDM models, structure formation is suppressed below the free-streaming scale, leading to a distinct cutoff in the power spectrum at this scale. 
This cutoff is represented in terms of the transfer function as follows:
\begin{equation}
    P_{\rm{WDM}}(k)=T^2(k)P_{\rm{CDM}}(k), 
    \label{Pwdm}
\end{equation}
where $P(k)$ is the power spectrum as a function of the comoving wavenumber, denoted by $k$. $T(k)$, is given by \citet{2001ApJ...556...93B}:
\begin{equation}
    T(k) = (1+(\alpha k)^{2\nu})^{-5/\nu}.
    \label{tf}
\end{equation}
In this study, the constant parameter $\nu$ is set to $\nu = 1.12$ as in \cite{viel2005} and \cite{viel2013}. The parameter $\alpha$ determines the position of the cutoff in the power spectrum. For instance, an increase in $\alpha$ results in a larger cutoff scale. This parameter is related to the mass of the thermal relic WDM particle, $m_{\rm{WDM}}$, and can be given as follows \citep{viel2005}:
\begin{equation}
    \begin{aligned}
        \alpha = \frac{0.049}{h\ \rm{Mpc}^{-1}}\left(\frac{m_{\rm{WDM}}}{1 \rm{keV}}\right)^{-1.15}\left(\frac{\Omega_{\rm{WDM}}}{0.4}\right)^{0.15} \\
        \times \left(\frac{h}{0.65}\right)^{1.3}\left(\frac{g_{\rm{WDM}}}{1.5}\right)^{-0.29},
        \label{alpha}
    \end{aligned}
\end{equation}
where $\Omega_{\rm{WDM}}$ is the density parameter contributed by WDM, and $g_{\rm{WDM}}$ is the degrees of freedom, which is set to $g_{\rm{WDM}}=1.5$. 

\begin{figure}[hbtp]
  \begin{center}
   \includegraphics[width=\columnwidth]{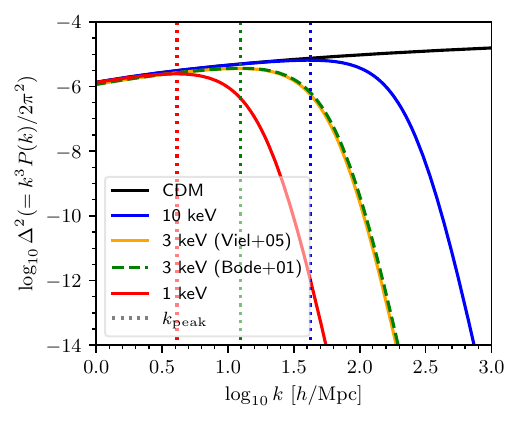} 
  \end{center}
 \caption{Dimensionless linear power spectra at $z=127$ used in the simulations.
 The horizontal axis represents the wavenumber $k$, while the vertical axis shows the dimensionless power spectrum on a logarithmic scale. The black line corresponds to the CDM model, the red line to the 1 keV WDM model, the orange and green lines to the 3 keV WDM model calculated from \cite{viel2005} and \cite{2001ApJ...556...93B}, respectively, and the blue line to the 10 keV WDM model. The dotted vertical lines indicate the wavenumbers, $k_{\rm{peak}}$, where the dimensionless power spectrum of each model reaches its maximum.

 {Alt text: Line plot with five colored curves representing dimensionless
power spectra for CDM and various WDM models. Vertical dotted lines
show the peaks of each spectrum. The plot uses logarithmic scales on
both axes.}
 }
 \label{fig:ps}
 \end{figure}

For the zoom-in simulations (1 keV, 10 keV, and CDM1), we generated initial conditions using the fitting formula for $\alpha$ derived by \citet{viel2005}. In contrast, for the uniform resolution box simulations (3 keV and CDM2), we adopted the expression from \citet{2001ApJ...556...93B}. These two expressions for $\alpha$ exhibit only minor differences. It should be noted that both the SASHIMI-W calculations and our analysis code consistently employ the $\alpha$ formula from \citet{viel2005} for all models. Consequently, in the 3 keV case, despite the initial conditions being generated using the \citet{2001ApJ...556...93B} formula, the simulation results are compared against SASHIMI-W predictions and analyzed based on the \citet{viel2005} definition. Importantly, this inconsistency has a negligible impact on our results. This is demonstrated in Fig.~\ref{fig:ps}, which presents a comparison of the power spectra calculated for the 3 keV WDM model using both $\alpha$ definitions.

In Fig.~\ref{fig:ps}, the horizontal axis represents the wavenumber on a logarithmic scale, $\log_{10} k$, and the vertical axis represents the dimensionless power spectrum, $\log_{10} \Delta^2$.
The black line corresponds to the CDM model, the red line to the 1 keV WDM model, the orange and green lines to the 3 keV WDM model calculated using the $\alpha$ expressions from \citet{viel2005} and \citet{2001ApJ...556...93B}, respectively, and the blue line to the 10 keV WDM model. Dotted lines indicate the wavenumber $k_{\rm{peak}}$ at which the dimensionless power spectrum of each model reaches its maximum.
As can be seen, the 3 keV power spectra computed using the two $\alpha$ expressions are almost indistinguishable, and their $k_\mathrm{peak}$ values coincide. Therefore, this inconsistency is not expected to impact our final conclusions.

The values of $\alpha$, $M_{\rm hm}$, and $k_{\rm peak}$ are listed in Table \ref{tab:modelparam}.
$M_{\rm hm}$ denotes the half-mode mass, which corresponds to the mass enclosed within a sphere of radius $\lambda_{\rm hm}/2$, where $\lambda_{\rm hm}$ is the half-mode wavelength \citep{2012MNRAS.424..684S}. The half-mode wavelength is defined as the scale at which the amplitude of the transfer function $T(k)$, given in Eq.~(\ref{tf}), decreases to $1/2$.
%Table 3
\begin{table}[h]
    \tbl{Values of $\alpha$, $M_{\rm hm}$, and $k_{\rm peak}$ for each model}{
        \begin{tabular}{@{}l|cccc@{}}
        \hline
        Model & $\alpha [{\rm Mpc}/h]$ & $M_{\rm hm}[{\rm M_{\odot}}/h]$ & $k_{\rm peak}[h/\rm Mpc]$\\
        \hline
        CDM & $0$ & $-$ & $-$\\
        1 keV & $0.048$ & $1.37\times 10^{10}$ & $4.10519$\\
        3 keV & $0.0141$ & $3.52\times 10^8$ & $12.3975$\\ 
        10 keV & $0.0037$ & $6.40\times 10^6$ & $42.3318$\\ 
        \hline
    \end{tabular}}\label{tab:modelparam}
\end{table}
%Figure 1

\subsection{Removal of spurious subhalos}\label{sec:removing}
In WDM models, as previously mentioned, structure formation is suppressed below the free-streaming scale. However, simulations of WDM models have shown that halos can still form below this scale (\cite{2007MNRAS.380...93W}). This occurs because numerical noise in the initial conditions can artificially induce structure formation, leading to the emergence of halos that are not physically meaningful. Therefore, before analyzing subhalos, it is crucial to identify and remove these spurious halos and subhalos from our dataset.
This section provides a brief overview of the method used to eliminate spurious halos.
%Figure 2
\begin{figure*}[hbtp]
  \begin{center}
   \includegraphics[width=\textwidth]{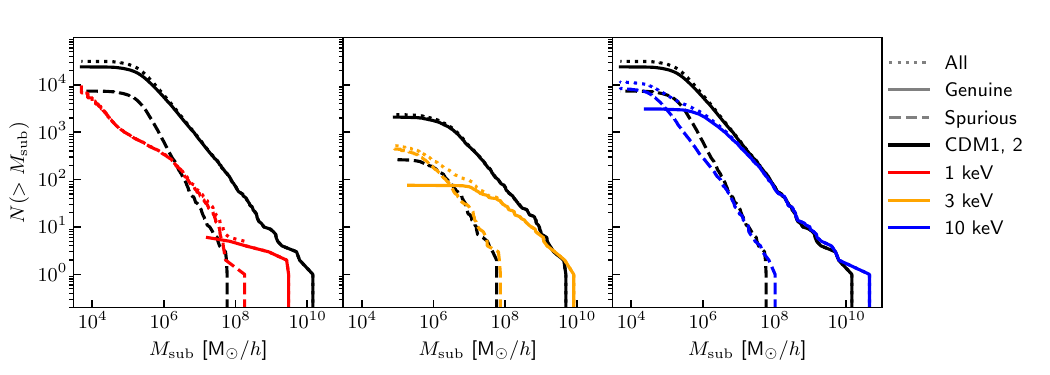} 
  \end{center}
 \caption{Cumulative mass functions for genuine subhalos and spurious subhalos at $z = 0$. 
 The horizontal axis represents the mass of subhalos, while the vertical axis shows the cumulative number of subhalos with a mass greater than a given value. The black line corresponds to CDM1 or CDM2, the red line to the 1 keV WDM model, the orange line to the 3 keV WDM model, and the blue line to the 10 keV WDM model. Solid, dashed, and dotted lines represent genuine, spurious, and all subhalos, respectively. For the 3 keV and CDM2 simulations, the plotted mass functions represent the mean values from five host halos.
 
 {Alt text: Line graph comparing cumulative mass functions of subhalos across multiple dark matter models. Different line styles indicate genuine, spurious, and total subhalo populations. Axes show subhalo mass and cumulative number.}
 }
 \label{fig:chmf1}
 \end{figure*}

The identification and removal of spurious halos are carried out from two distinct perspectives. 
The first criterion examines the shape of the initial Lagrangian region of halos (proto-halos). As noted by \citet{Lovell_2014}, proto-halos that originate from the gravitational collapse of physical initial density fluctuations (genuine halos) tend to be relatively spherical, whereas those of spurious halos appear more flattened.
To quantify this, they introduce sphericity, denoted by $s$, 
which is defined in terms of the eigenvalues $\lambda_1$, $\lambda_2$, and $\lambda_3$ ($\lambda_1 \geq \lambda_2 \geq \lambda_3$) of the particle distribution tensor $M_{ij}$ within the proto-halo:
\begin{equation}
s = \sqrt{\lambda_3/\lambda_1}. 
\end{equation}
The particle distribution tensor $M_{ij}$ is defined as
\begin{equation}
M_{ij} = \sum_p m_p {x_p}_i {x_p}_j,
\label{Mij}
\end{equation}
where the sum runs over all particles in the proto-halo, $m_p$ is the mass of particle $p$, and ${x_p}_i$ represents the $i$-th component ($i = 1,2,3$) of the particle’s position relative to the proto-halo’s center of mass.

Since subhalos can lose their mass due to tidal effects after merging with the main halo, we perform calculations using the halo’s state prior to merging. Specifically, $s$ is computed using the initial positions of particles in the main progenitor of a subhalo at the time when it first reached half of its maximum mass, denoted as $M_{\rm half}$, during its evolutionary history.
Following \citep{Lovell_2014}, a threshold, $s_{\rm cut}$ is defined  as the 1st percentile of $s$ in the corresponding CDM simulation. 
Halos with $s$ below this threshold are classified as spurious and excluded from further analysis.

The second criterion for identifying spurious halos is based on mass. \citet{2007MNRAS.380...93W} found that the threshold mass, denoted as $M_{\rm lim}$, below which spurious halos dominate the subhalo mass function depends on both the simulation resolution and the cutoff scale of the power spectrum. They define $M_{\rm lim}$ as
\begin{equation}
M_{\rm lim} = 10.1\Bar{\rho} d k_{\rm peak},
\label{Mlim}
\end{equation}
where $\Bar{\rho}$ is the mean density of the universe, $d$ is the mean inter-particle separation, and $k_{\rm peak}$ is the wavenumber at which the dimensionless power spectrum reaches its maximum.
However, some genuine halos may have masses below $M_{\rm lim}$, while some spurious halos may exceed this threshold. Thus, simply removing all subhalos with masses below $M_{\rm lim}$ could lead to the loss of real halos or the retention of spurious ones. To address this, \citet{Lovell_2014} introduced a refined mass threshold, $M_{\rm min}$, which is empirically determined by comparing simulations with different resolutions. It is given by
\begin{equation}
M_{\rm min} = \kappa M_{\rm lim},
\label{Mmin}
\end{equation}
where $\kappa$ is chosen such that the number of subhalos with $M > M_{\rm min}$ and $s > s_\mathrm{cut}$ remains consistent with the number of genuine subhalos identified by comparison with higher-resolution simulations.
%Table 4
\begin{table*}[h]
    \tbl{Mass $M_{\rm main}$ and virial radius $R_{\rm vir}$ of the main halos at $z=0$}{
    \begin{tabular}{@{}cl|ccccc@{}}
        \hline
        & Model & (1) & (2) & (3) & (4) & (5)\\
        \hline
        \multirow{5}{*}{$M_{\rm main}[\rm M_{\odot}]$} & 1 keV & $9.567\times10^{11}$ & $-$ & $-$ & $-$ & $-$\\ 
        & 10 keV & $9.145\times10^{11}$ & $-$ & $-$ & $-$ & $-$\\
        & CDM1 & $9.205\times10^{11}$ & $-$ & $-$ & $-$ & $-$\\
        & 3 keV & $2.151\times10^{12}$ & $7.074\times10^{11}$ & $5.155\times10^{11}$ & $5.753\times10^{11}$ & $5.136\times10^{11}$\\
        & CDM2 & $2.183\times10^{12}$ & $7.657\times10^{11}$ & $5.586\times10^{11}$ & $5.558\times10^{11}$ & $4.987\times10^{11}$\\
        \hline
        \multirow{5}{*}{$R_{\rm vir}[{\rm kpc}/h]$} & 1 keV & $2.00\times10^2$ & $-$ & $-$ & $-$ & $-$\\
        & 10 keV & $1.97\times10^2$ & $-$ & $-$ & $-$ & $-$\\
        & CDM1 & $1.98\times10^2$ & $-$ & $-$ & $-$ & $-$\\ 
        & 3 keV & $2.62\times10^2$ & $1.81\times10^2$ & $1.63\times10^2$ & $1.63\times10^2$ & $1.69\times10^2$\\
        & CDM2 & $2.64\times10^2$ & $1.86\times10^2$ & $1.67\times10^2$ & $1.67\times10^2$ & $1.61\times10^2$\\
        \hline
    \end{tabular}}\label{tab:main}
\end{table*}

In this study, we adopt the values $s_{\rm cut} = 0.16$ and $\kappa = 0.5$, as specified by \citet{Lovell_2014}. The corresponding values of $M_{\rm{min}}$ are $1.1 \times 10^8 \,\rm M_{\odot}$ for the 1 keV model, $4.5 \times 10^7 \,\rm M_{\odot}$ for the 3 keV model, and $1.0 \times 10^6 \,\rm M_{\odot}$ for the 10 keV model. Using these thresholds, we will identify and remove spurious halos to obtain the subhalo mass functions for each model. For consistency, we also exclude CDM subhalos with $s < s_\mathrm{cut}$ from our analysis.

\subsection{The semi-analytical model}
Here, we briefly summarize SASHIMI-W, the semi-analytical model employed in this work.
At present ($z=0$), subhalos are characterized by their mass $m$, the scale radius $r_s$ and characteristic density $\rho_s$ of the Navarro-Frenk-White (NFW) profile (\cite{1997ApJ...490..493N}), and the truncation radius $r_t$, which defines the point where the subhalo density sharply drops to zero (\cite{2008MNRAS.391.1685S}).

However, subhalos undergo tidal stripping after being accreted into a main halo, leading to modifications of their internal structure both before and after accretion. When the truncation radius $r_t$ falls below $0.77$ times the scale radius $r_s$, significant tidal mass loss occurs, eventually resulting in the complete disruption of the subhalo (\cite{2003ApJ...584..541H}). Given the complexity of subhalo evolution, it is crucial to develop an evolutionary model that connects their present-day properties (e.g., $m$, $r_s$, $\rho_s$, $r_t$) with their pre-accretion values. 

SASHIMI-W provides a framework for modeling these processes in the context of WDM cosmology \citep{2016MNRAS.460.1214L, 2018PhRvD..97l3002H, 2022PhRvD.106l3026D}.  
To this end, it incorporates several key modifications to its parent CDM version. In particular, SASHIMI-W accounts for the distinctive properties of WDM halos by adopting a modified mean concentration--mass--redshift relation, based on the work of \cite{2016MNRAS.460.1214L}, which reproduces the characteristic turnover in concentration at the truncation scale. Halo concentrations are assigned by sampling from a log-normal distribution centered on the predicted mean for a given mass and redshift. The half-mode mass, $M_{\rm hm}$, is not treated as a free parameter; instead, it is derived directly from the input WDM power spectrum, which is determined by the WDM particle mass (see Table~\ref{tab:modelparam}), and sets the characteristic scale for the suppression of structure formation.

\section{Result}\label{result}
We focus on the main halos whose virial masses, $M_\mathrm{main}$, are similar to the Milky Way halo mass $\sim  10^{12} \, \rm M_\odot$. In the zoom simulations, we chose a halo with $M_\mathrm{main}$ very close to $10^{12} \, \rm M_\odot$. On the other hand, in the uniform box simulations, $M_\mathrm{main}$ ranges from $\sim 5 \times 10^{11}$ to $\sim 2 \times 10^{12} \, \rm M_\odot$. Table \ref{tab:main} presents the mass, $M_{\rm main}$, and the virial radius, $R_{\rm vir}$, of the main halos at $z=0$.

The results of removing spurious halos using the method described in Section \ref{sec:removing} are shown as the cumulative mass function of subhalos at $z=0$ in Fig.~\ref{fig:chmf1}.
In this figure, the horizontal axis represents the subhalo mass as $M_{\rm sub}$, and the vertical axis represents the number of subhalos with a mass greater than a certain value as $N(>M_{\rm sub})$. The black line corresponds to CDM1 or CDM2, the red line to 1 keV, the orange line to 3 keV, and the blue line to 10 keV WDM. Solid, dashed, and dotted lines represent genuine, spurious, and all subhalos, respectively. 
Overall, low-mass subhalos in the WDM models are predominantly spurious, particularly in the warmer WDM models. 
We also find that the resolution of our simulations is sufficiently high that for masses where the WDM subhalo mass functions deviate from those of the CDM, the genuine halos dominate, with little influence from the spurious halos.

\begin{figure*}[hbtp]
    \begin{center}
    \includegraphics[width=\textwidth]{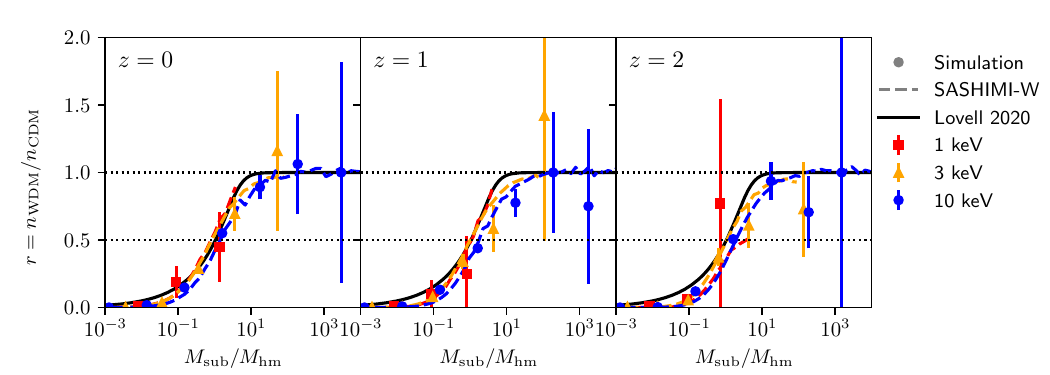}
    \end{center}
    \caption{Ratios, $r$, of the subhalo differential mass functions in WDM models relative to the CDM model. The horizontal axis shows subhalo mass normalized by the half-mode mass, while the vertical axis shows the ratio of WDM to CDM differential mass functions. The solid black line represents the fitting function from \citet{2020ApJ...897..147L}, while other colors follow the same scheme as in Fig.~\ref{fig:chmf1}. Symbols denote simulation results, and dashed lines show predictions from SASHIMI-W. The panels correspond to redshifts $z=0$ (left), $z=1$ (center), and $z=2$ (right).
    
    {Alt text: Three-panel line and symbol plot showing the ratio of warm dark matter to cold dark matter subhalo mass functions at three redshifts. Each panel compares simulation results, semi-analytical model, and the fitting function as functions of normalized subhalo mass.}
    }
    \label{fig:rhmf}
\end{figure*}

In Fig.~\ref{fig:rhmf}, we present the ratios, $r$, of the differential mass functions predicted by SASHIMI-W for the WDM models to that for the CDM model. We define the CDM model as the case where the WDM particle mass is set to $10^6$ keV in SASHIMI-W\footnote{Since the CDM version of SASHIMI, SASHIMI-C \citep{sashimi-c}, employs a different window function from SASHIMI-W, we utilize SASHIMI-W with a very heavy particle mass as a CDM version of SASHIMI for consistency.}. For comparison with SASHIMI-W predictions, we show the ratios $r$ obtained from our simulations. Additionally, we present the fitting function $r_\mathrm{fit}$ by \citet{2020ApJ...897..147L}:
\begin{equation}
    r_{\rm fit} = \dfrac{n_{\rm WDM}}{n_{\rm CDM}} = \left[1+\left(\dfrac{4.2M_{\rm hm}}{M_{\rm sub}}\right)^{2.5}\right]^{-0.2}.
    \label{rfit}
\end{equation}
%Figure 3

In this figure, the horizontal axis shows subhalo mass normalized by the half-mode mass, while the vertical axis displays the ratio of the WDM to CDM subhalo differential mass functions. Colors follow the same scheme as in Fig.~\ref{fig:chmf1}. The solid black line represents the fitting function $r_{\rm fit}$ proposed by \citet{2020ApJ...897..147L}, symbols indicate simulation results, and dashed lines show SASHIMI-W predictions. For the 3 keV and CDM2 simulations, we plot the mean value of $r$ across the five main halos. Errors are calculated as:
\begin{equation}
    r_{\rm err} = \sqrt{\left(\dfrac{1}{n_{\rm CDM}}n_{\rm err}^\mathrm{WDM}\right)^2+\left(\dfrac{n_{\rm WDM}}{{n_{\rm CDM}}^2}n_{\rm err}^\mathrm{CDM}\right)^2},
    \label{rerr}
\end{equation}
where 
\begin{equation}
    n_{\rm err}^\mathrm{X} = \dfrac{n_{\rm X}}{\sqrt{\mathrm{d}N_{\rm X}}}.
    \label{nerr}
\end{equation}

We find that the ratios $r$ predicted by SASHIMI-W for the 1 keV, 3 keV, and 10 keV WDM models align well with our simulation results. The simulations reveal a slight redshift evolution of $r$, which SASHIMI-W accurately captures. The fitting function becomes progressively less accurate at higher redshifts. Even at $z = 0$, SASHIMI-W predictions show better agreement with our simulations than the fitting function.

\section{Discussion}\label{discussion}
Our analysis demonstrates that SASHIMI-W accurately reproduces the ratio $r$ between WDM and CDM differential mass functions across all tested WDM particle masses and redshifts. This indicates that once the CDM differential mass function is established, the corresponding WDM differential mass function can be reliably derived. Nevertheless, a more stringent validation requires assessing whether SASHIMI-W can directly predict the individual differential mass functions themselves, rather than solely their ratios.

\begin{figure*}[hbtp]
    \begin{center}
    \includegraphics[width=\textwidth]{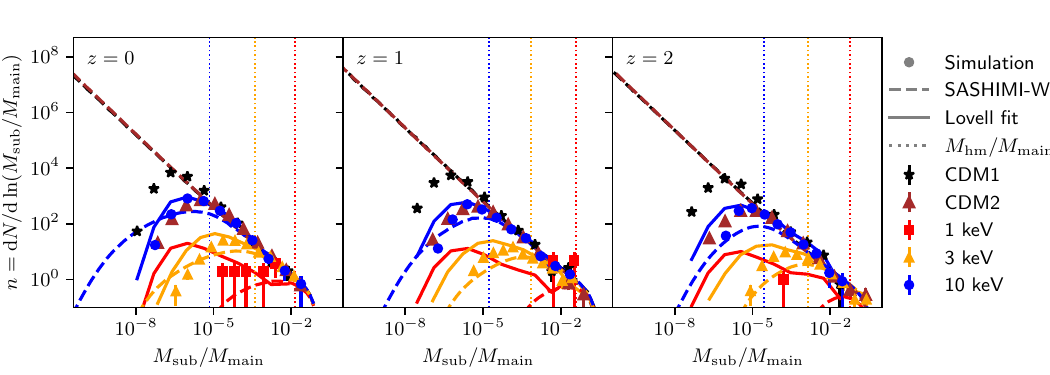}
    \end{center}
    \caption{ 
    Differential subhalo mass functions from simulations and SASHIMI-W. Color coding denotes the models: black for CDM1, brown for CDM2, red for 1 keV WDM, orange for 3 keV WDM, and blue for 10 keV WDM. Symbols represent simulation data, while dashed lines show the predictions from SASHIMI-W. Solid lines indicate the differential mass function given by the fitting formula of \citet{2020ApJ...897..147L}. For the 3 keV WDM model, the mass functions are averaged over the five main halos. The left panel corresponds to $z = 0$, the center panel to $z = 1$, and the right panel to $z = 2$. The horizontal axis shows the subhalo mass normalized by the main halo mass, while the vertical axis shows the number of subhalos per mass interval. Vertical dotted lines indicate the values of $M_{\rm hm}/M_{\rm main}$ for each WDM model. 
    
    {Alt text: Three-panel plot comparing subhalo mass functions from simulations and semi-analytical model, and those derived from the fitting function at three redshifts. Each panel shows multiple curves and symbols as functions of normalized subhalo mass.}
    }
    \label{fig:dhmf}
\end{figure*}

Fig.~\ref{fig:dhmf} shows the differential subhalo mass functions for both CDM and WDM models, as measured from our simulations, and compares them with the predictions from SASHIMI-W and those derived from the fitting function of \citet{2020ApJ...897..147L}. The line styles and symbols follow those used in Fig.~\ref{fig:rhmf}, while the color scheme matches that of Fig.~\ref{fig:chmf1}. The horizontal axis indicates the subhalo mass normalized by the main halo mass, and the vertical axis shows the number density of subhalos per logarithmic mass bin. Error bars are calculated as described in Eq.~(\ref{nerr}). Vertical dotted lines mark the value of $M_{\rm hm}/M_{\rm main}$ for each WDM model.

For CDM, the differential subhalo mass functions derived from SASHIMI-W slightly underestimate the number of medium- and low-mass subhalos with mass ratios $M_{\mathrm{sub}}/M_{\mathrm{main}} < 10^{-3}$. This discrepancy may be attributed to halo-to-halo variations inherent in the simulations. Nevertheless, SASHIMI-W generally exhibits good agreement with the simulation results. Notably, even within CDM simulations, a low-mass cutoff is observed, mirroring the behavior seen in WDM models. This cutoff is not a physical feature but rather a numerical artifact arising from the limited resolution of the simulations. Specifically, it results from the absence of initial density fluctuations below the Nyquist wavelength imposed by the initial conditions.

For the 1 keV WDM model, the differential mass functions predicted by SASHIMI-W slightly underestimate the simulation results at the peak at $z = 0$, although they remain within the error bars in other regions. Given the limited number of halos in this model, these results are statistically less robust compared to those for the 3 keV, 10 keV, and CDM cases.

The differential mass functions for the 3 keV and 10 keV WDM models calculated by SASHIMI-W exhibit mass-dependent characteristics. For the 3 keV WDM model, SASHIMI-W accurately reproduces simulation results for high-mass subhalos where $M_{\mathrm{sub}}/M_{\mathrm{main}} > 10^{-3}$ across all redshifts. At $z = 0$, it underestimates the number of medium-mass subhalos in the range $10^{-6} < M_{\mathrm{sub}}/M_{\mathrm{main}} < 10^{-3}$, while overestimating the number of low-mass subhalos with $M_{\mathrm{sub}}/M_{\mathrm{main}} < 10^{-6}$. At $z = 1$ and $z = 2$, SASHIMI-W consistently underestimates the number of both medium- and low-mass subhalos compared with the simulations. For the 10 keV WDM model, SASHIMI-W accurately reproduces the simulation results for high-mass subhalos across all redshifts. However, at $z = 0$ and $z = 1$, SASHIMI-W underestimates the number of medium-mass subhalos while overestimating the number of low-mass subhalos. Conversely, at $z = 2$, SASHIMI-W underestimates the number of both medium- and low-mass subhalos compared to the simulations. These discrepancies are not evident in Fig.~\ref{fig:rhmf} due to the linear scaling of its vertical axis. 
%Figure 4

As an additional point of comparison, the differential mass function predicted by the \citet{2020ApJ...897..147L} fitting function shows good agreement with the simulation results in the high-mass regime. It also appears to match the data better than SASHIMI-W in the intermediate-mass range, where SASHIMI-W exhibits systematic underestimation. However, this apparent superior performance is misleading, as the fitting function is constructed by multiplying a fitting ratio by the baseline mass function from the corresponding CDM simulation, $n_{\rm CDM}$. In this mass range, SASHIMI-W underpredicts even the CDM baseline, which exaggerates the discrepancy in its WDM prediction when compared directly to the fitting function. As demonstrated in Fig.~\ref{fig:rhmf}, when examining the ratio of WDM to CDM mass functions, SASHIMI-W consistently reproduces the simulation results with higher fidelity than the fitting function across all mass scales.

Several factors may contribute to the overestimation of low-mass subhalo counts by SASHIMI-W compared to the simulations. Firstly, insufficient resolution in the simulations could artificially suppress the number of subhalos, as observed in the CDM simulations. Secondly, the tidal effects incorporated in SASHIMI-W might be inadequately modeled. Lastly, our criteria for removing spurious (sub)halos may be too stringent for low-mass subhalos.

Conversely, we hypothesize that the underestimation of subhalo counts by SASHIMI-W at $M_\mathrm{sub}/M_\mathrm{main} \sim 10^{-5}$ may be attributed to one of two factors: either the tidal effects in SASHIMI-W (which are appropriately calibrated for CDM models) act too strongly in WDM models at this mass range, or there is insufficient removal of spurious subhalos from the simulations within this mass range.

Given the uncertainties in the removal of spurious subhalos, we believe that SASHIMI-W is sufficiently accurate for our purposes. For more rigorous tests, future work would benefit from WDM simulations that are inherently free from spurious halos \citep{coldice, hahn2016}.

It is essential to assess whether the accuracy of SASHIMI-W, as evaluated in this study, is sufficient for placing observational constraints on the WDM particle mass. Constraints based on Milky Way satellite counts primarily probe the subhalo mass function in the range $10^{-5} \lesssim M_{\rm sub}/M_{\rm main} \lesssim 10^{-3}$, which corresponds to subhalos hosting ultra-faint to classical dwarf galaxies (e.g., \cite{2021PhRvL.126i1101N}). Our analysis (Fig.~\ref{fig:dhmf}) indicates that SASHIMI-W systematically underpredicts subhalo abundances in this mass range compared to our simulation results.

This underprediction, at the level of $\sim 0.2$--0.3 dex, could bias WDM mass constraints toward higher particle masses—that is, artificially colder models may be required to reproduce the observed satellite population. Recent studies (e.g., \cite{2022PhRvD.106l3026D}) emphasize that theoretical predictions must reach an accuracy of $\sim 10$--20\% in this regime to enable robust constraints. In contrast, SASHIMI-W shows discrepancies by factors of $\sim 2$–3 in some parts of this mass range.

Nevertheless, the ratio of the WDM to CDM subhalo mass functions predicted by SASHIMI-W is in significantly better agreement with the simulations. This suggests that, given a reliable CDM baseline, SASHIMI-W remains a viable tool for deriving meaningful WDM constraints—--potentially eliminating the need for computationally expensive, high-resolution simulations of massive WDM particles.

Stellar stream gaps offer a complementary probe of the lower end of the subhalo mass spectrum, $10^{-6} \lesssim M_{\rm sub}/M_{\rm main} \lesssim 10^{-4}$, extending into the regime of dark, starless subhalos that lie beyond the reach of satellite counts (e.g., \cite{2019ApJ...880...38B}). In this regime, SASHIMI-W exhibits more complex behavior. In the upper portion ($10^{-5} \lesssim M_{\rm sub}/M_{\rm main} \lesssim 10^{-4}$), it continues to underpredict subhalo abundances, potentially leading to overly stringent constraints if observed stream gaps are predominantly caused by subhalos in this mass range—again favoring colder WDM models to compensate for the shortfall.

At even lower masses ($M_{\rm sub}/M_{\rm main} \lesssim 10^{-5}$), however, SASHIMI-W begins to overpredict subhalo counts. This may result in overly permissive constraints, erroneously favoring lighter WDM particles, since the model suggests that even warmer WDM candidates can produce sufficient gap-forming subhalos. It is worth noting that this overprediction may stem from resolution effects in the simulations---specifically, the simulations may underpredict subhalo abundances at this mass scale due to limited numerical resolution.

Finally, a robust interpretation of stream gaps must account for contamination from non-dark-matter sources, such as internal stream dynamics, baryonic perturbers (e.g., giant molecular clouds), and uncertainties in the Galactic potential. These systematic uncertainties highlight the need for improved modeling of tidal disruption in WDM cosmologies and for more stringent criteria to identify and exclude spurious subhalos before SASHIMI-W can be confidently used for precision cosmological inference.

\section{Conclusion}
To validate the semi-analytical model and fitting function for subhalo mass functions in WDM models, we conducted high-resolution cosmological $N$-body simulations for both CDM and WDM models. Our analysis compared results from these simulations with predictions from SASHIMI-W, a semi-analytical framework designed to infer subhalo properties in WDM models, and the fitting function developed by \citet{2020ApJ...897..147L}.

SASHIMI-W accurately reproduced our simulation results regarding the ratio $r$ of the differential mass functions between WDM and CDM models, showing consistency across different WDM particle masses and redshifts. Notably, our simulations and the SASHIMI-W model revealed a slight redshift dependence in the ratio $r$.

Furthermore, we investigated whether the differential mass functions predicted by SASHIMI-W align with our simulation results in detail. For the CDM model, SASHIMI-W reproduced the overall simulation outcomes well, with only minor discrepancies, which could be within the halo-to-halo variation. In contrast, for the WDM model, SASHIMI-W showed good agreement with simulation results only for high-mass subhalos, while exhibiting deviations in the medium- and low-mass regimes, which are not evident in the linear scaling of $r$. These discrepancies likely stem from either inaccuracies in SASHIMI-W's treatment of tidal stripping effects or uncertainties in removal of spurious subhalos in our simulations. 

Our findings suggest that when the subhalo mass function for a CDM model is well-established, one can derive the corresponding WDM subhalo mass function using SASHIMI-W with slight uncertainties. 
Since the ratio $r$ slightly depends on both WDM particle mass and redshift, we recommend using SASHIMI-W over the fitting function. 

\begin{ack}
This study is supported by JSPS/MEXT KAKENHI Grants (JP21H04496, JP20H05861, 25H00671), JST SPRING, Grant Number JPMJSP2119, and MEXT as ‘Program for Promoting Researches on the Supercomputer Fugaku’ (Toward a unified view of the Universe: from large-scale structures to planets, Grant No. JPMXP1020200109). S.A. has been supported by JSPS/MEXT KAKENHI Grants (20H05850, 24K07039). T.I. has been supported by IAAR Research Support Program in Chiba University Japan, MEXT/JSPS KAKENHI (Grant Number JP23H04002), and JICFuS. Numerical computations and analysis were carried out on Cray XC50 and computers at the Center for Computational Astrophysics, National Astronomical Observatory of Japan.
\end{ack}

\appendix %%%%%%%%%%%%%%%%%%%%%%%%%%%%%%%%%%%%%%%%%%%%%%%%%%%%%%%%
\section{Results for Individual halos for the CDM2 and the 3 keV model}\label{appendix:individual_halos}
 %Figure 5
\begin{figure*}[hbtp]
  \begin{center}
   \includegraphics[width=\textwidth]{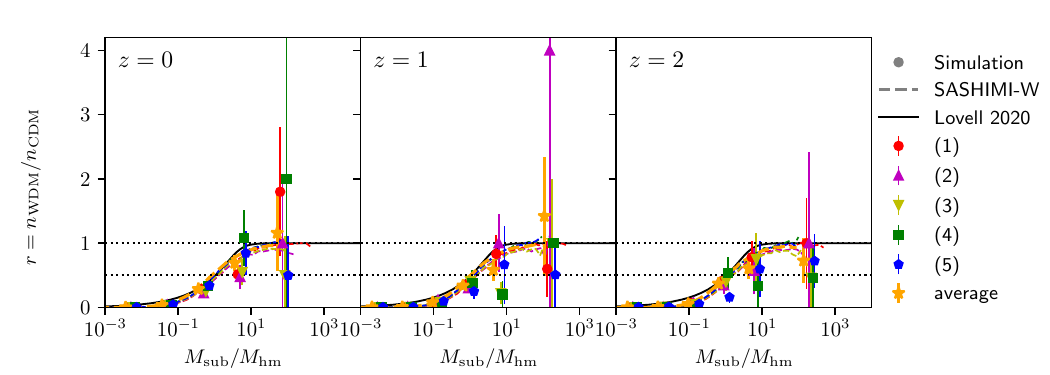} 
  \end{center}
 \caption{Ratios of subhalo differential mass functions, $r$, for the individual halos in the 3 keV WDM model at redshifts $z=0,\ 1,$ and $2$ (from left to right).
 The horizontal axis and the vertical axis are the same as in Fig.~\ref{fig:rhmf}. Each color represents the ratio for an individual main halo: red for (1), magenta for (2), yellow for (3), green for (4), and blue for (5). The orange shows the average of the five, which is also plotted in Fig.~\ref{fig:rhmf}. The solid black line represents the fitting function from \citet{2020ApJ...897..147L}. Symbols denote simulation results, and dashed lines show predictions from SASHIMI-W. Note that the data points for each halo are slightly displaced horizontally to prevent their error bars from overlapping.
 {Alt text: A three-panel line and symbol plot showing the ratio of 3 keV warm dark matter to cold dark matter subhalo mass functions for five individual halos and average on these halos at three redshifts. Each panel compares simulation results, semi-analytical model, and the fitting function as function of normalized subhalo mass.}
 }
 \label{fig:rhmf_3keV}
 \end{figure*}

In this appendix, we present the detailed results for the CDM2 and 3\,keV WDM models, which were discussed in Secs.~\ref{result} and \ref{discussion} in terms of averaged quantities.

Figure~\ref{fig:rhmf_3keV} shows the halo-to-halo variation in the ratio of subhalo mass functions, $r$, for the 3\,keV WDM model. In the regime $M_{\mathrm{sub}}/M_{\mathrm{hm}} \lesssim 10$, the scatter among the five halos is sufficiently small. At higher mass ratios, $M_{\mathrm{sub}}/M_{\mathrm{hm}} \gtrsim 10$, the scatter increases slightly but remains largely within the error bars of the average. This indicates that the most massive host halo in our sample, halo~(1), does not significantly bias the average values presented in Sec.~\ref{result}.

\begin{figure*}[t]
  \begin{center}
   \includegraphics[width=\textwidth]{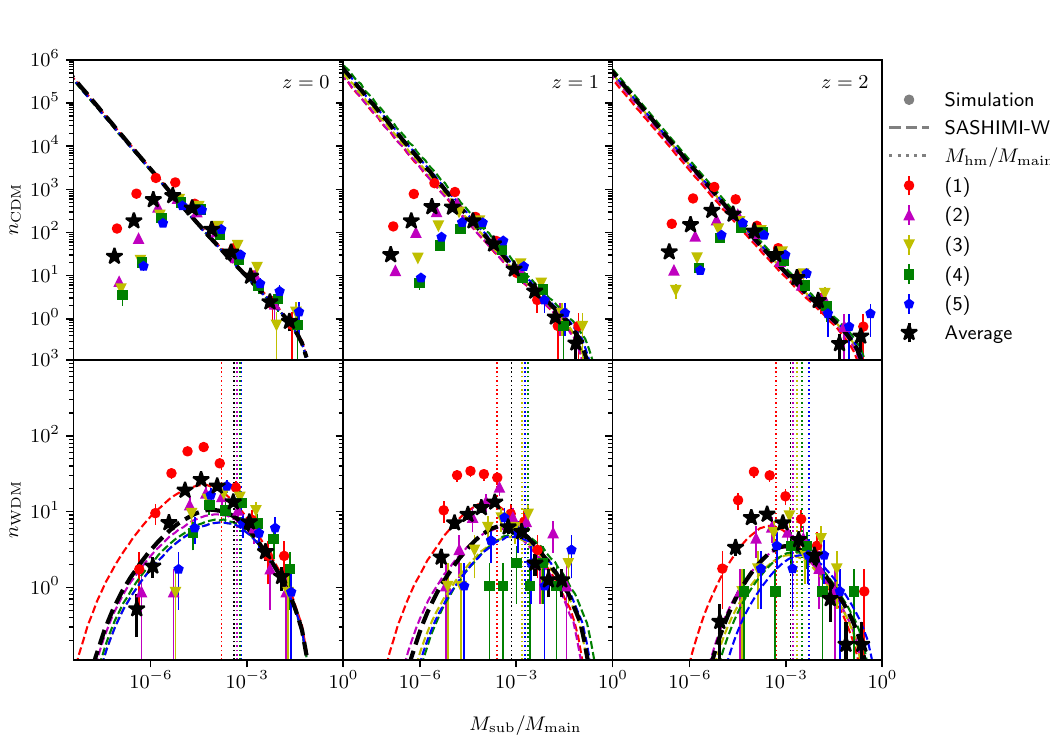} 
  \end{center}
 \caption{Differential mass functions of subhalos for the individual halos in the CDM2 and the 3 keV WDM model at redshifts $z=0,\ 1,$ and $2$ (from left to right).
 The upper panels show results for the CDM2 model, while the lower panels show those for the 3 keV WDM model. The horizontal axis and the vertical axis are the same as in Fig.~\ref{fig:dhmf}. The colored vertical dotted lines indicate the value of $M_{\rm hm}/M_{\rm main}$ for each corresponding main halo, while the black vertical dotted line marks the average value. Each color represents the differential mass function for an individual main halo: red for (1), magenta for (2), yellow for (3), green for (4), and blue for (5). The black data show the average of the five, which is also plotted in Fig.~\ref{fig:dhmf}. Symbols denote simulation results, and dashed lines show predictions from SASHIMI-W. Note that the data points for each halo are slightly displaced horizontally to prevent their error bars from overlapping. 
 {Alt text: A six-panel grid of plots comparing simulated and modeled subhalo mass functions for five individual halos and average on these halos at three redshfits. The top row is for a cold dark matter model and the bottom row is for a 3 keV warm dark matter model. Each panel shows multiple curces and symbols as functions of normalized subhalo mass.}
 }
 \label{fig:dhmf_3keV}
 \end{figure*}

Figure~\ref{fig:dhmf_3keV} displays the differential subhalo mass functions, $n$, for individual halos in the CDM2 and 3\,keV WDM models. Each trend is broadly consistent with the averaged results discussed in Sec.~\ref{discussion}. Notably, the peak of the mass function, $n_{\rm peak}$, shifts depending on the mass of the host halo---a mass-dependent trend that is qualitatively reproduced by SASHIMI-W. However, despite capturing this trend, SASHIMI-W systematically underestimates the value of $n_{\rm peak}$ across all halos when compared to the simulation results.
%Figure 6

% Any journal's BST file (e.g., apj.bst) can be used as PASJ's BST is unavailable.    
\bibliographystyle{aasjournal}
\bibliography{reference}
% \bibliographystyle{****}
% \bibliography{****}

\end{document}